\documentclass[12pt,preprint,useAMS]{aastex}

\newcommand{\eb}{\begin{equation}}
\newcommand{\ee}{\end{equation}}

\shorttitle{Zonal astrometric errors}
\shortauthors{Makarov et al.}
\begin{document}

\title{Correlated and zonal errors of global astrometric missions: a spherical harmonic solution}
\author{V.V. Makarov, B.N. Dorland, R.A. Gaume, G.S. Hennessy, C.T. Berghea, R.P. Dudik}
\affil{United States Naval Observatory, Washington, DC 20392, USA}
\email {vvm@usno.navy.mil}
\and

\author{H.R. Schmitt}
\affil{Computational Physics, Inc., Springfield, VA 22151, USA}
\email {}
\begin{abstract}
We propose a computer-efficient and accurate method of estimation of spatially correlated 
errors in astrometric
positions, parallaxes and proper motions obtained by space and ground-based astrometry missions. In our
method, the simulated observational equations are set up and solved for the coefficients of
scalar and vector spherical harmonics representing the output errors, rather than
for individual objects in the output catalog. Both accidental and systematic correlated errors of astrometric
parameters can be accurately estimated. The method is demonstrated on the
example of the JMAPS mission, but can be used for other projects of space astrometry,
such as SIM or JASMINE.
\end{abstract}

\keywords{astrometry --- parallaxes --- proper motions --- methods: numerical}

              
\def\C0{$c_0$ }
\def\M0{$m_0$ }

\section{Introduction}
 
Projects of global space astrometry, such as Hipparcos \citep{esa}, Gaia \citep{per}, SIM \citep{unw}
or JMAPS \citep{gau}, result in very large systems of observational equations, which
contain the unknown parameters of interest, as well as an even larger
number of instrument calibration and satellite attitude parameters.
The structure of this system depends on the architecture of the mission,
adopted schedule of observations and properties of the instrument. In its most general matrix form, 
a linearized observational equation of space astrometry is
\eb
\boldmath{A\,a}+\boldmath{C\,c}+\boldmath{P\,p}=\bf{y},
\label{obs.eq}
\ee
where the three main types of unknowns are separated, namely,
the astrometric parameters $\bf{a}$, the calibration
parameters $\bf{c}$ and the attitude parameters $\bf{p}$. The
measurements collected during the on-orbit operation are involved in the
right-hand side $\bf y$ of the observational equations. These measurements include accidental
and systematic errors. The accidental error arises from a completely
stochastic, unpredictable process and occurs spontaneously in each observation, independent of the
previous or subsequent observations, for example, photon shot noise in
the incoming light signal. The systematic error has a certain deterministic cause,
such as a particular state of the instrument, and may in principle be avoided or
mitigated by setting adequate requirements to the mission design and the instrument.
The problem we are going to consider is how
to calculate the propagation of accidental and systematic errors in the astrometric
solution in the general case. In particular, we are concerned about the
occurrence of correlated and zonal errors in the resulting proper motions and parallaxes.

We are considering in this paper the position-correlated part of the total absolute
error of an astrometric catalog or reference frame.  It can be visualized as
a smooth pattern in the distribution of the absolute ``observed minus true" error
on the celestial sphere. The actual distribution of absolute error is sampled on
a discrete set of objects (e.g., stars and quasars), so that the uncorrelated part will look like
discrete noise, but if a smoothing procedure is applied to this sampled
function, the underlying large-scale pattern will emerge. This smooth pattern
will be stochastic, i.e., unpredictable, if it is caused by the accidental
errors of the measurements, or deterministic, and therefore predictable, if
it is caused by a systematic measurement error. In reality, any catalog produced
by a space mission will include a mixture of accidental and systematic
errors. The goal of a well-designed astrometric mission is to keep the systematic error
smaller that the accidental error. However, an external catalog of superior accuracy is
usually required to verify that this goal is achieved. For example, a comparison of the
Hipparcos catalog with the FK5 system of 1535 reference stars revealed zonal errors in
the latter that were larger than the estimated accidental error per star \citep{mign}. The type
of large-scale errors considered in this paper is different from the accumulated small-scale
errors, which originate in the shared attitude error for stars observed within the same
field of view (or the same scan for a scanning mission) \citep{vl}. The details of 
such accumulation of small-scale error in the in-scan direction for scanning missions is 
highly dependent on specific mission implementation and must be modeled separately on a case-by-case basis.

The smoothness of spatially correlated large-scale errors invokes the use of spherical orthogonal
functions for their representation and analysis. A parallax error is a scalar
function of spherical coordinates, and is best described by the scalar spherical
harmonics. Proper motion and position errors are tangential vector fields on the
unit sphere, and should be properly represented by vector spherical harmonics.
The spherical harmonics are basis functions, therefore, any smooth function of
coordinates can be accurately decomposed into a linear combination of a sufficient
number of spherical harmonics. Some of the spherical harmonic terms correspond 
to specific physical parameters. 
For example, the zero-order scalar spherical harmonic of
parallax is a constant function (unity) on the sky. This term is the well-known 
parallax zero-point. The zero-point is the average
of all absolute parallax errors of a given catalog, which is different from zero.
This common type of error is of special significance in astrometry,
because it directly affects the cosmic distance scale based on trigonometric
parallaxes. A common offset of parallaxes is relatively more important for distant
objects, hence, the empirical calibration of such mostly distant calibrators as Cepheids or RR Lyr-type stars is especially vulnerable to a zero-point error. All higher-order harmonics do
average to zero on the sphere and have more localized effect. As another example of coupling
between physical parameters and specific large-scale errors, the Oort's parameters describing the
differential rotation of the Galaxy are represented by certain vector spherical harmonics
of the proper motion field \citep{mamu}.

One might naively assume that random uncorrelated noise in the observational data results in
a uniform distribution of error among the different types of zonal error, i.e., in a flat spherical harmonic
spectrum. In fact, this is never the case, as the mode of operation,
the geometrical order of observation, and the finite field of view of the telescope, all
lead to a strongly non-uniform propagation of observational noise in different orders of
correlated error. The technique presented in this paper goes back to the idea of using
orthogonal functions for the analysis of one-dimensional ``abscissae" errors of
the Hipparcos mission \citep[for definition of abscissae, see][Vol. 2]{esa}. The concept of Hipparcos was built around a stable and self-calibrating
\textsl{basic angle} separating the two viewing directions of the telescope. \citet{hoy}
suggested that the problem of propagating perturbations of the basic angle becomes
tractable if these perturbations are represented by a Fourier series. Using this idea, \citet{mak1}
showed that Fourier time-harmonics of the basic angle propagate with
uneven magnification coefficients into the corresponding harmonics of star abscissae, with the $60\deg$-harmonic
being the dominating one in the output. The reason for this peculiar propagation of
white noise is the value of the basic angle, which is close to $60\deg$. The technique was
further developed and applied to the R{\o}mer project\footnote{The R{\o}mer concept was the precursor of the Gaia mission.} by \citet{mak2}. The propagation of periodic
perturbations of star abscissae into specific spherical harmonics of absolute error is a more
complicated issue, which depends on the scanning law and the geometry of reference great
circles. This generalization was implemented  for a Hipparcos-like design by \citet{mak3}, where
the 2D errors of parallax were represented by spherical harmonics. Using an orthogonal
basis of functions on the unit sphere puts the concept of {\it absolute} astrometry on a rigorous mathematical
footing. The proposed JMAPS and SIM missions, unlike Hipparcos and GAIA, employ only one
viewing direction \citep{zac}. The emergence of correlated errors in these projects is defined by
the density of overlapping observational frames and the accuracy of quasar constraints.

\section{Observational equations and quasar constraints}

Linearized observational equations of global astrometry relate
astrometric parameters of interest, satellite attitude parameters and
instrument calibration parameters to astrometric observable parameters.
Although the actual relations between these parameters are nonlinear,
the required linearization can be achieved by taking the perturbation form
and limiting the Taylor expansions on the left-hand side of the equations to first
order. If the initial guess or prior knowledge of the fitting
parameters is close to the truth, this first-order approximation provides an
accurate solution; otherwise, the linearization and global
solutions have to be iterated.

A common property of all proposed space astrometry projects is that most of the
nuisance parameters (e.g., calibration and attitude) 
entering the observational equations should be determined from
the same observations along with the star parameters. Essentially, an instrument for space
astrometry is self-calibrating and self-navigating. Some of the crucial parameters,
e.g., the basic angle for Hipparcos, the baseline length for SIM, are stable
by engineering requirements; they should be re-determined relatively infrequently
during the mission.
On the contrary, the attitude parameters are unique for each astrometric frame or
scan, and therefore, generate the bulk of nuisance parameters. The Euler angles (or quaternions) of
spacecraft attitude are approximately known from the navigation system, including
a separate star tracker device, but much more accurate values of these parameters are determined
from the main observations themselves \citep{lim}. 

It was shown on the example of the SIM project that coupling between the attitude
unknowns and the astrometric unknowns can cause a loss of condition and a
non-uniform propagation of errors in a global solution \citep{mami}. A strict
relation exists between the basis vectors of parallax distribution, obtained by
the singular value decomposition (SVD)
of the corresponding part of condition equations, and the scalar harmonics sampled
on a discrete set of stars. The reciprocal singular values are simply the coefficients
of different degrees and orders of error, propagating into the final parallax solution. By virtue of the relatively
small size of the design matrix, the SIM grid solution was ideally suited for rigorous
mathematical analysis of various aspects of error propagation. Other astrometric missions
invoke much larger Least Squares (LS) problems, and SVD analysis becomes intractable. In this paper, 
we are setting out to develop a numerical method to estimate the propagation
of large-scale correlated errors in very large LS solutions.

In the perturbation form, the unknowns in Eq. \ref{obs.eq} are small
corrections to {\it a priori} parameters describing the stars and the state of the instrument,
and $\bf{y}$ is the vector of small differences between the the predicted and the actual measurements.
The grand design matrix can be constructed, in principle, from the individual blocks $\bf{A}$,
$\bf{C}$ and $\bf{P}$, although it is never done in practice because of its huge size.
The standard method of solving such problems is {\it block adjustment} \citep[e.g.][]{von}, 
using the natural
sparsity and structure of the design matrix. Briefly, there have been two algorithms
considered for large astrometric problems, the {\it iterative} block adjustment and the
global {\it direct} solution \citep{buc}. Hipparcos, the only implemented astrometric space mission
thus far, relied on the iterative adjustment, in which the major blocks of unknown
parameters were estimated and updated in turns, while keeping the other types of
parameters fixed, resulting in a number of iterations across the range of mission
parameters. The convergence of the large-scale iterations can not be taken for granted, but
should be verified by dedicated simulations.
A similar algorithm of iterative block adjustment has been developed for the
Gaia data analysis system \citep{lam}.
On the mathematical grounds, a global adjustment, which is a simultaneous,
one-step solution for the multitude of mission parameters, should be more exact, faster
and easier to analyze, but it poses a considerable implementation challenge for huge LS problems
with a large number of nonzeros.

The block structure of the grand design matrix is defined by two different types of dependancies
of the unknowns. The astrometric unknowns ${\bf a}$ are object-dependent, i.e., they comprise
independent sets of several unknowns for each object. Five astrometric unknowns per star are usually
considered, namely, position components, parallax and proper motion components. The calibration unknowns
${\bf c}$ and the attitude ${\bf p}$ are mostly, but not exclusively, time-dependent. If the observational
equations are sorted by time, the nonzero elements of the design matrix corresponding to the attitude unknowns
are found in tight, relatively small blocks, because the attitude of an astrometric instrument
is fast-varying. The calibration parameters can be discretized too, so that a separate set of calibration
unknowns is assigned to a fixed interval of observations, which we will call a calibration block
in this paper. The calibration parameters are expected to change slowly with time. Therefore, the
corresponding blocks of nonzero elements are longer that the attitude blocks. It is convenient to
adjust the discretization steps in such a way that the boundaries of the blocks are aligned, so that
an integer number of attitude blocks corresponds to each calibration block. The astrometric unknowns
in such a design matrix are scattered across its entire length, because the same object is observed
multiple times during the mission.

It is sometimes practical to eliminate the attitude and calibration unknowns in the equations
rather than solve for them directly along with the astrometric unknowns. This elimination
is achieved by the QR factorization of each block and the subsequent ${\bf Q}^T$ orthogonal transformation of the
remainder of the design matrix and the right hand side of the equations, as described in
\citep{mami}. Because of the nested structure of the blocks, the elimination procedure becomes
hierarchical, the smaller attitude blocks eliminated first, followed by the calibration blocks.
As a result, the number of unknowns is significantly reduced. However, this reduction comes at a cost,
because the design matrix becomes much denser. Obviously, nonzero off-diagonal elements are generated
for any pair of objects, which were observed within the same calibration block. The degree of
densification depends on the average number of objects within a calibration block. To avoid
intractably dense matrices, smaller calibration partitions are preferred. In the JMAPS global
solution, several large-scale calibrations parameters are solved for each frame, along with the
three attitude unknowns. In that case, the direct LS solution is obtained for about 29 million
unknowns with 1 million grid objects, or 34 million unknowns with 2 million grid objects.
After the QR elimination, only 5 or 10 million unknowns remain, respectively, but the design
matrix is much denser. The number of equations to be solved is 144 or 288 million, respectively.
Our idea presented in this paper is that in many cases, it is sufficient to consider the correlated errors of the simulated mission, rather than the individual errors of numerous grid objects. This strategy helps
to reduce the number of astrometric unknowns to manageable levels, fully capturing an important
characteristic of mission performance. The mathematical technique is described in \S\S \ref{sh.sec} and
\ref{sol.sec}.

\subsection{Spherical harmonics}
\label{sh.sec}
The astrometric part of observational equations can be written as
\eb
\left( \delta{\bf s_0}+\delta{\boldmath \mu}(t-t_0)+\delta\varpi (({\bf s}_0\cdot {\bf b})\,{\bf s}_0-{\bf b})\right) \cdot 
{\bf\tau}= y.
\label{astro.eq}
\ee
where $\delta{\bf s_0}$ and  $\delta{\bf \mu}$ are the unknown corrections to mean position and mean proper motion at $t_0$, tangential to the celestial sphere at ${\bf s_0}$, which is the assumed position unit vector of the
objects at time $t_0$, $\delta\varpi$ is the unknown correction to parallax, and ${\bf b}$ is
the position vector of the space craft with respect to the barycenter of the Solar system at
the time of observation $t$, assumed to be known. The vector ${\bf\tau}$ is a certain fiducial
direction defined by the instrument, for example, the baseline vector of SIM, or the nominal
scanning direction of Gaia. This vector depends on the instantaneous attitude and the calibration
parameters, but here it is assumed to be known, because all the nuisance parameters have been
separated in the linearized equations into independent blocks. For a two-dimensional pointing mission 
such as JMAPS, two condition equations emerge from a single observation, since there are two fiducial
directions in the focal plane, corresponding to the rows and columns of the detector array. The right-hand
side of Eq.~\ref{astro.eq} includes the measurement and additive accidental and systematic errors.

\begin{figure}[here!]
\epsscale{1}
\plotone{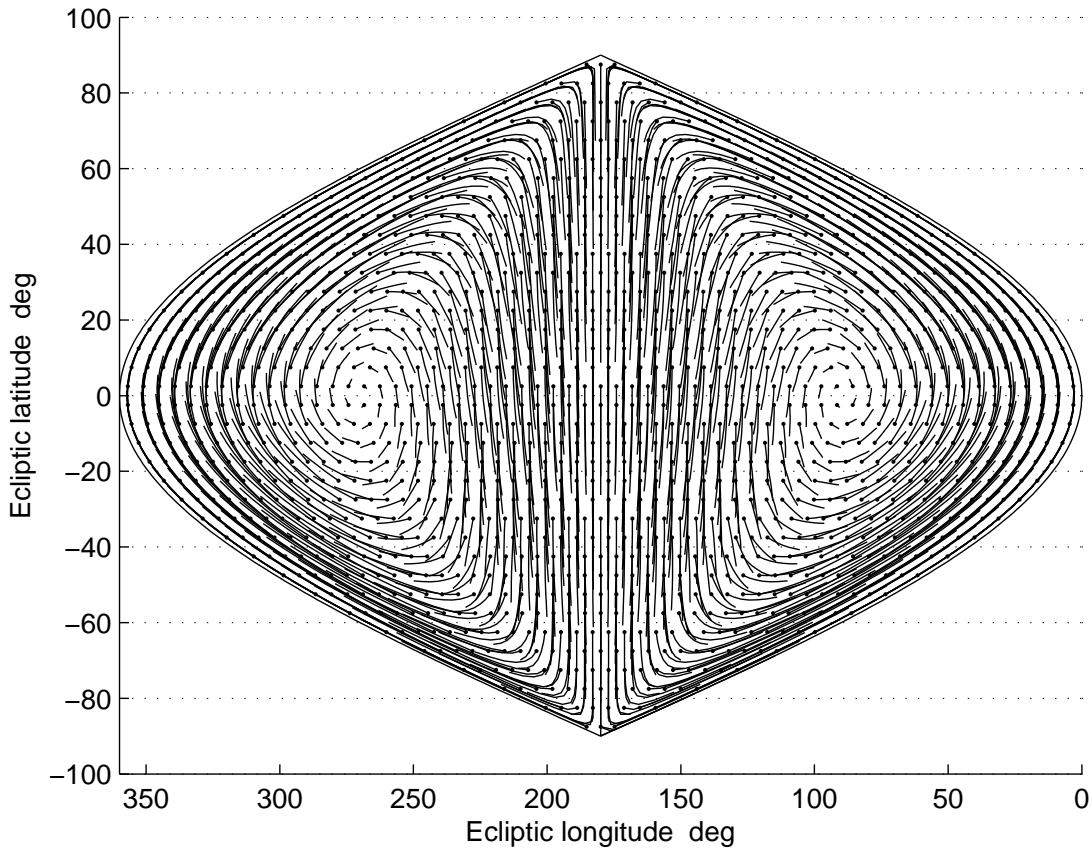}
\caption{The magnetic vector spherical harmonic ${\bf H}_1^{-1}$.}
\label{v6.fig}
\end{figure}

The astrometric condition equations are linear and can be solved by direct LS with or without elimination
of the nuisance parameters. The main technical problem arises from the size of the normal matrix,
which require supercomputing facilities and advanced algorithms. For Gaia, the size is so large that
a direct LS solution is deemed impossible, and the adopted iterative solution still takes a long time
\citep{omu}. Solving directly for up to 34 million unknowns has been proven feasible with a
specially adapted PARDISO solver (PARallel DIrect SOlver, part of the Intel Math Kernel Library), 
but it still takes several hours of computing time to complete.
For the testing and verification purposes, full-scale runs of the global solution have to be performed
multiple times, with various input data. Our idea presented in this paper is that in many cases, it is
sufficient to consider the distribution of error on the sphere, rather than individual errors of numerous
grid objects. Thus, we substitute the object-dependent astrometric unknowns in Eq.~\ref{astro.eq}
with the expansions in spherical harmonics, which are functions of celestial coordinates, e.g., the ecliptic
coordinates $\lambda$ and $\beta$:
\begin{eqnarray}
\delta {\bf s}_0 &=& \sum_{j=1}^{\infty}S_j{\bf V}(\lambda,\beta) \hspace{12mm}{\rm position}\nonumber \\
\delta {\bf \mu}_0 &=& \sum_{j=1}^{\infty}M_j{\bf V}_j(\lambda,\beta)  \hspace{9mm}{\rm proper~motion}\nonumber \\
\delta  \varpi_0 &=& \sum_{j=1}^{\infty}Q_j\,U_j(\lambda,\beta)  \hspace{10mm}{\rm parallax}
\label{ser.eq}
\end{eqnarray}
with $U_j$ being the scalar spherical harmonics and ${\bf V}_j$ the vector spherical harmonics. For a detailed
description of spherical harmonics see, e.g., \citep{mamu}. Here we only reproduce some basic formulae.
The vector harmonics ${\bf V}_j$ are composed of two types of functions, called magnetic and
electric harmonics, $\bf H_n^m$ and $\bf E_n^m$ respectively. These
vector harmonics are derived via partial derivatives of the scalar
spherical harmonics over angular coordinates, viz.:

\begin{eqnarray}
{\bf H}_n^m(\lambda,\beta)&=& \left[ \frac{\partial U_n^m(\lambda,\beta)}{\partial \beta}
\vec \tau_\lambda-\frac{1}{\cos \beta}\frac{\partial U_n^m(\lambda,\beta)}{\partial \lambda}
\vec{\tau_\beta}\right]\nonumber\\
{\bf E}_n^m(\lambda,\beta)&=& \left[ \frac{1}{\cos \beta}\frac{\partial U_n^m(\lambda,\beta)}
{\partial \lambda}\vec{\tau_\lambda}+\frac{\partial U_n^m(\lambda,\beta)}{\partial \beta}
\vec \tau_\beta \right]
\end{eqnarray}
The pair of vectors $\{\tau_\lambda,\tau_\beta\}$ define the local tangential coordinate
system in the plane orthogonal to ${\bf s}_0$, directed north and east, respectively.
Spherical harmonics $U_n^m$ are counted by degrees $n=0,1,\ldots$ and orders $m=-n,-n+1,\ldots,n$. Explicitly,
\begin{eqnarray}
\label{sph.eq}
U_n^m &=& \sqrt{\frac{2n+1}{2\pi}\frac{(n-m)!}{(n+m)!}}     P_n^{m}(\sin \beta)\cos m\lambda,    \hspace{8mm} m>0,\nonumber\\
      &=& \sqrt{\frac{2n+1}{4\pi}}                          P_n^{0}(\sin \beta)               \hspace{40mm} m=0,\nonumber\\
      &=& \sqrt{\frac{2n+1}{2\pi}\frac{(n-|m|)!}{(n+|m|)!}} P_n^{|m|}(\sin \beta)\sin |m|\lambda, \hspace{4mm} m<0
\end{eqnarray}
where $P_n^m$ are the associated Legendre polynomials. The first pair of
vector harmonics are generated from the scalar harmonic $U_1^0$, with the
electric component $\cos \beta \vec{\tau}_\beta$ and the magnetic component
$\cos \beta \vec{\tau}_\lambda$. The common index $j$ used in Eq.~\ref{ser.eq}, introduced for simplicity,
counts all individual harmonics in the following manner: for each degree $n$ all orders of electric
harmonics are lined up, followed by all orders of magnetic harmonics. A particular vector harmonic,
${\bf V}_6$, which is the magnetic harmonic ${\bf H}_1^{-1}$ is depicted in Fig.~\ref{v6.fig}. It is
equivalent to a left-handed rotation around the pole at $\lambda=90\degr$, $\beta=0$.
 
\subsection{A direct solution on a laptop}
\label{sol.sec}
The most important advantage of the spherical harmonics is that they are orthogonal on the unit sphere
in the space of continuous functions, or nearly orthogonal when discretized on a large set of uniformly
distributed points. In the latter case, the deviation from orthogonality is negligibly small for a sufficiently
large number of grid stars ($\ga 10^4$) of uniform density on the sky. The degree of uniformity and the
number of grid stars should be higher if the higher order harmonics are to be used in the direct
global solution. Normally, the lower orders of spherical harmonics are of interest, where the largest
correlated errors emerge. Therefore, the series in Eq.~\ref{ser.eq} can be truncated in the new
condition equations for the fitting coefficients $S_j$, $M_j$, and $Q_j$. The solution
for a subset of model terms is exact if the terms are orthogonal. In practice, the degree of orthogonality
should be verified for a given distribution of stars and weights (in a weighted LS). Rearranging the
discretized spherical harmonics for the three types of astrometric unknowns as columns of a design
matrix, the unknown coefficients and the right-hand side data as column vectors, the LS problem can be
written in this compact matrix form:
\eb
\left[{\bf V}_s \, {\bf V}_\mu\, {\bf U}_\varpi \right]\;\left[\begin{array}{c}
{\bf S}\\{\bf M}\\{\bf Q}\end{array}\right]={\bf y}
\label{mat.eq}
\ee
The length of the design matrix is still very large in this setup, because it includes all
the observations of grid stars. For JMAPS, it is about 144 million for 1 million grid stars. The width of the design matrix, on the other hand, is defined
by how many spherical harmonic terms we want to solve for.  Indeed, a sufficiently accurate solution
can be obtained for any subset of spherical harmonics, as long as the columns of the design matrix
are nearly orthogonal. This is verified by computing the correlation coefficients of the covariance matrix.
They should all be small, e.g., less than 0.01 in absolute value. If this is the case, including more
terms in the design matrix will not change the solution for $S_j$, $M_j$, and $Q_j$ significantly.
The number of unknowns can be made comfortably small for a given  computer. We found, for
example, that a global solution can be obtained within 1 hour for 400 unknowns on a regular laptop
computer.

Even with a limited number of unknowns, the design matrix is too large to be handled in fast memory
without swapping with disk. However, there is no need to keep the entire matrix in memory if the
observations are sorted by time. If the design matrix ${\bf D}=\left[{\bf V}_s \, {\bf V}_\mu\, {\bf U}_\varpi \right]$ is divided into a number of blocks ${\bf D}_j$ in the vertical dimension (not necessarily of the
same length), the normal matrix is the sum of the normal sub-matrices, ${\bf D}^T{\bf D}=\sum_j {\bf D}_j^T
{\bf D}_j$. The accumulated normal matrix can be easily inverted due to its relatively small size, resulting
in the covariance matrix, ${\bf Cov}=\left({\bf D}^T{\bf D}\right)^{-1}$. The off-diagonal elements of ${\bf Cov}$ should be small due to the near-orthogonality of the model terms, unless some additional global parameters
are included. The diagonal elements are the variances of unit weight of the coefficients of spherical
harmonics $S_j$, $M_j$, and $Q_j$. If the observations are weighted by the expected standard deviation of
measurement error, the variances are the squares of the standard errors carried by the corresponding spherical
harmonics. The total mission-average variance is approximately the sum of the variances of the complete set
of harmonics for each of the 5 astrometric parameters. Since we obtain the variances for a limited set of
harmonics, the total mission-average error can not be inferred from this computation. However, the uncertainty
of specific harmonic components is accurately computed. 

\section{Results and Discussion}
JMAPS is a pointing astrometric telescope with a single viewing direction. Without the ability of Hipparcos
to simultaneously observe stars that are far apart on the sky, the required rigidity of the reference system
and the accuracy of astrometric parameters is achieved through measuring a number of carefully selected
ICRF and radio-mute quasars and other extragalactic objects. These objects provide absolute constraints
on positions (using the ICRF coordinates of superior accuracy), parallaxes and proper motions, which are
negligibly small because of the extreme remoteness. The entire sky is observed with a 4-fold overlap.
Astrometric observations are normally made around the great circle perpendicular to the sun direction.
Some 72 observations per object are expected to be collected in three years. For the simulations
described in this paper, we used a catalog of 44 ICRF quasars and 80 compact extragalactic sources
(QSO), which are not in the ICRF. All these reference quasars are brighter than magnitude
15. Only a subset of all observable stars, usually between 1 and 2 million
strong, is used in the direct global solution. 

The attitude unknowns are represented by three parameters for each frame, viz., the translations along
the axes of the detector and rotation around the boresight vector. These unknowns are eliminated frame
by frame, reducing the number of conditions by three. The instrument calibration unknowns are
represented in these simulations by sets of up to 28 Zernike polynomials of field coordinates, separately for
either coordinate in the detector plane. The first Zernike polynomials, which are constant, are
excluded to avoid deficiency of rank, because they are indistinguishable from the attitude translations.
In our simplified simulations, the calibration parameters are assumed to be constant within calibration
blocks of equal length. Usually, blocks of 92 or 96 consecutive frames are used, corresponding to roughly
50 min of uninterrupted observations. As soon as a complete calibration block is collected, the QR
factorization is applied, and the remaining astrometric equations are pre-multiplied with ${\bf Q}^T$,
as well as the right-hand side. The number of condition equations is further reduced by $2\,N_Z$ for
each calibration block, with $N_Z$ being the number of calibration parameters. This algorithm allows
us to include a set of {\it global} parameters, which do not vary with time, such as the PPN $\gamma$-parameter. The accuracy or precision of global parameters can be reliably estimated, because they
are mostly correlated with the low-order components. The number of vector spherical unknowns is
$2(N_V+1)^2-2$, where $N_V$ is the limiting degree, and the number of scalar spherical harmonics
(for parallax) is $(N_V+1)^2$.

\subsection{Accidental errors}
\begin{figure}[here!]
\epsscale{1.15}
\plottwo{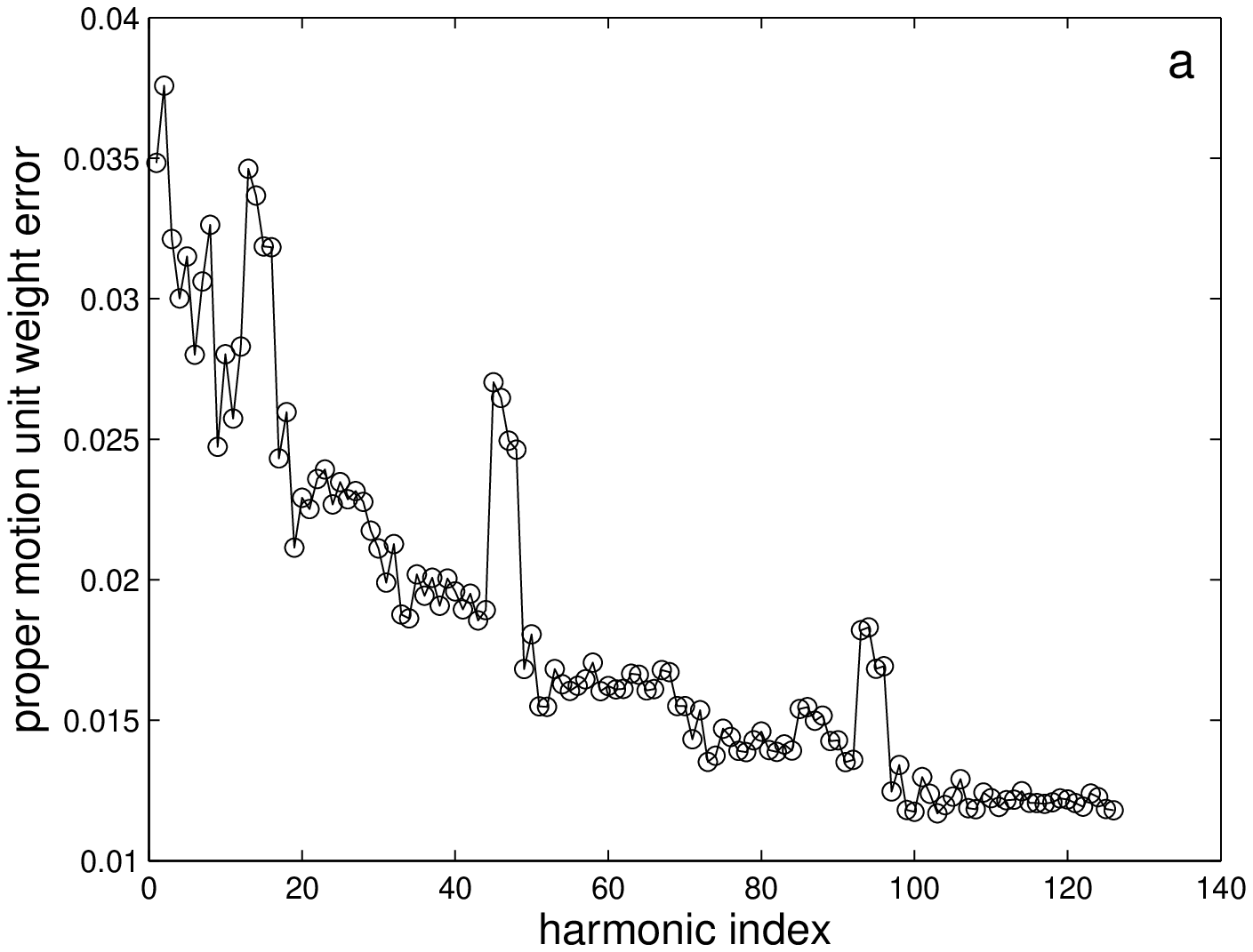}{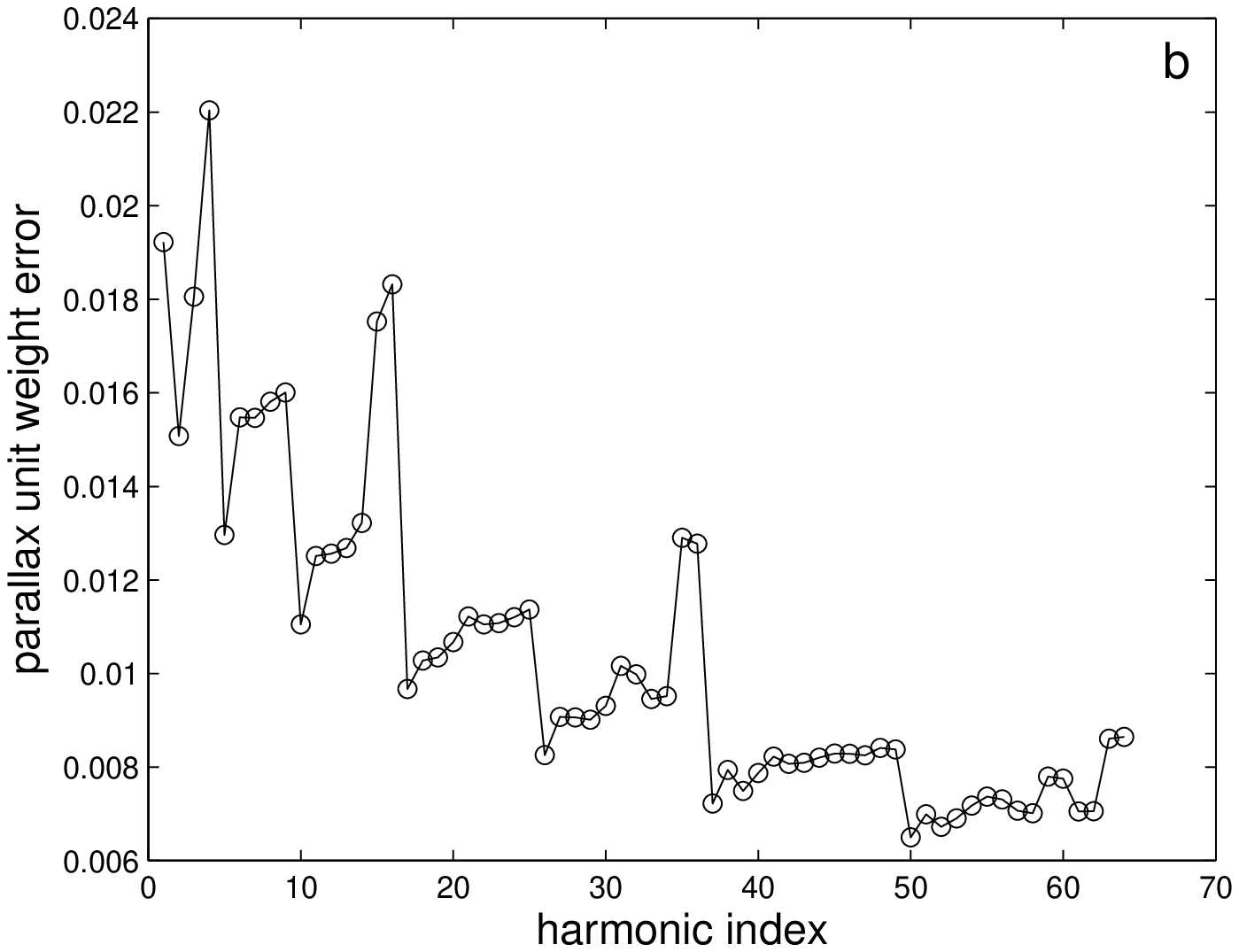}
\caption{Standard deviations of the coefficients of spherical harmonics representing
accidental error in a) proper motions, and b) parallax. The errors are normalized to unit weight.
 } \label{sig.fig}
\end{figure}

The statistical properties of accidental correlated errors are defined by the global covariance matrix of
the coefficients of spherical harmonics.
The diagonal of the covariance matrix at $N_V=7$ includes 126 vector spherical harmonic coefficients
for positions and proper motions each and 64 scalar spherical harmonic coefficients for parallax.
The square roots of the portions of the diagonal corresponding to each astrometric parameter
are the standard deviations of error of unit weight, represented by a particular harmonic.
For example, the standard deviation of the parallax zero-point error is the the standard deviation of
the first spherical harmonic coefficient multiplied by the weighted average single measurement precision
of stars and quasars.
Figs. \ref{sig.fig} a and b show the standard deviations of harmonic errors of proper motions and parallax,
respectively, obtained from a typical simulation of JMAPS mission. Generally, we find that the correlated
errors in all three parameters fairly rapidly decline with the degree of spherical harmonic. To use an analogy
from spectroscopy, in that
sense, the spectrum of accidental errors is ``red". There are some obvious ``spectral lines", however,
which are caused by the observing pattern and the distribution of reference quasars on the sky.

\begin{figure}[here!]
\epsscale{1}
\plotone{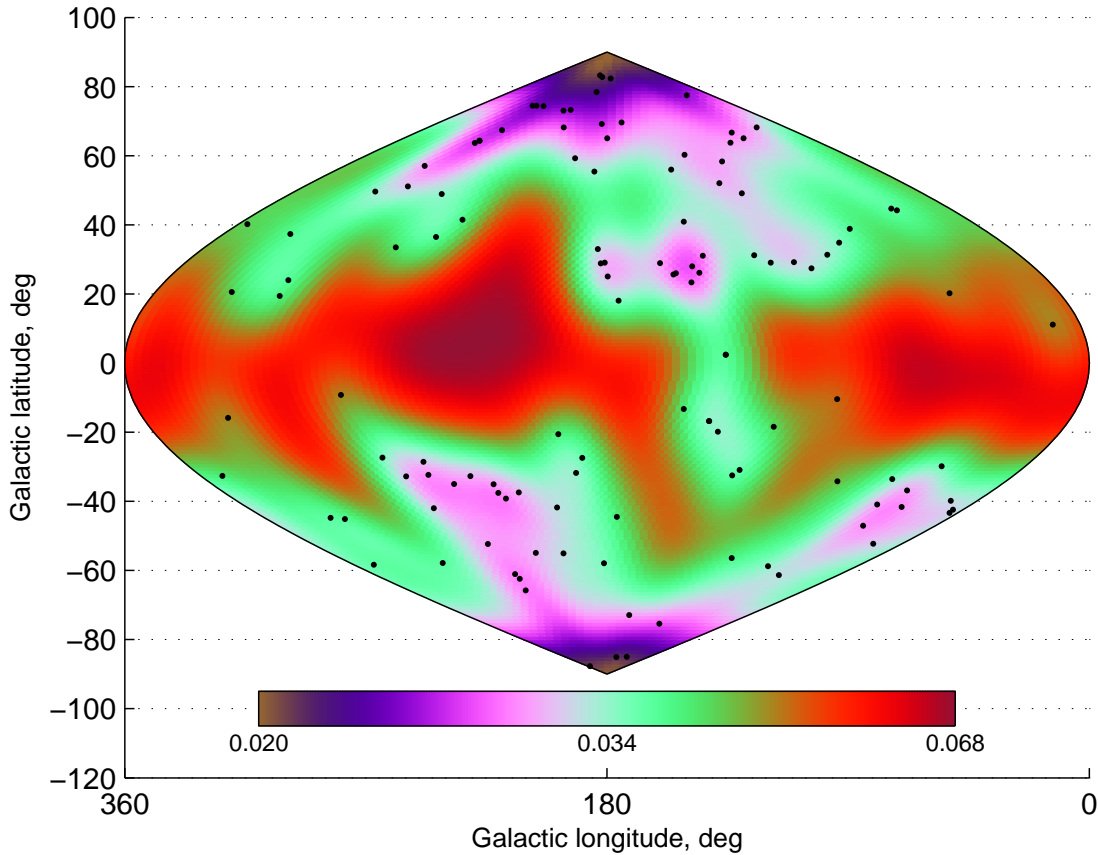}
\caption{Sky distribution of the standard deviation of parallax unit weight error for a simulated JMAPS
mission. The reference quasars used to constrain the parallax solution are marked with black dots.}
\label{par.fig}
\end{figure}

We find that the distribution of accidental error becomes flatter with a significantly larger number
of grid quasars. The relative height of the ``spectral lines" depends on the distribution of
grid quasars on the sky and the composition of the calibration model. Large holes in the distribution
of quasars cause considerable degradation of the overall performance. Using the near-orthogonality of
the discretized spherical function, the variance of accidental error of, e.g., proper motion at a given point $(\lambda,
\beta)$ can be estimated as
\eb
{\rm var}[{\bf \mu}(\lambda,\beta)]\simeq \sum_j {\rm var}\left[M_j\right]\|{\bf V}_j(\lambda,\beta)\|^2
=\sum_j C_{\mu,jj}\|{\bf V}_j(\lambda,\beta)\|^2
\label{var.eq}
\ee
where ${\bf C}_{\mu}$ is the corresponding part of the covariance matrix. Since this decomposition is limited
to a finite set of spherical functions, only a lower bound of the total error can be obtained.
Still, the distribution of the error carried by the lower-order harmonics is very informative.
For example, one can estimate the degree of inhomogeneity of the correlated error on the celestial
sphere, which can be significant for JMAPS. Fig. \ref{par.fig} depicts the distribution of the standard
deviation of the total accidental error of parallax, which is contained in the first 64 scalar spherical
harmonics. The plot is rotated into the Galactic coordinate system to emphasize the impact of
the zone of avoidance around the Galactic plane where quasars brighter than 15 mag can not be found.
The quasars, which were used to constrain the global solution for parallax, are shown as black dots.
The build-up of error in the areas devoid of grid quasars is quite obvious. As a way of improving
the overall mission performance, fainter quasars should be found near the Galactic plane and included
in the grid.
 
It should be noted that the distribution in Fig. \ref{par.fig} corresponds to the {\it expectancy}
of the spatially correlated error rather than to the outcome of a single mission. In other words, it shows what would
emerge if the same mission is simulated many times with random noise in the input data, and
the sample variance of the resulting errors is computed at each point. A single random realization of
correlated error can be computed by
\eb
\delta{\bf \mu}(\lambda,\beta) = \left({\bf C}_\mu^\frac{1}{2} {\bf r}\right)^T {\bf V}(\lambda,\beta)
\label{del.eq}
\ee
with ${\bf C}_\mu^\frac{1}{2}$ being the unique positive definite matrix square root of ${\bf C}_\mu$,
${\bf r}$ a random vector drawn from ${\cal N}(0,1/N_V)$ of $N_V$ elements, and ${\bf V}(\lambda,\beta)$
the column vector of the values ${\bf V}_j(\lambda,\beta)$. These transformations are performed separately
for each coordinate direction for the vector-valued parameters (position and proper motion).

\subsection{Systematic errors}
Systematic errors of global solutions are much harder to predict and analyze, because there are multiple
sources of such errors, which are rarely known  beforehand. Slowly varying perturbations of observational
data, caused by external circumstances, are of special interest, as they can bring about smooth,
large-scale errors. The orientation of the astrometric satellite with respect to the sun direction
is one of the conceivable sources of systematic error. The angle between the sun direction, which is confined to the ecliptic plane, and the viewing direction changes in a predetermined way, because the entire celestial sphere should be observed as uniformly as possible. The thermal flow inevitably changes inside the
telescope, resulting in slowly varying instrument parameters, e.g., the effective focal length or the basic
angle for Gaia. If these variations are correlated with the celestial coordinates, there is no averaging
out of the perturbation, and the error can propagate into the final catalog. In many cases, such specific
physical influences can not be accurately modeled or predicted. A more general modeling approach can be
exploited, where a certain perturbation is represented as a set of basis functions. For example, a systematic
variation of the basic angle can be represented as a Fourier series of the sun angle, and each of the
Fourier terms can be simulated separately. The previous studies for Hipparcos and SIM indicate that many
of such elementary perturbations are benign, in that they cause a relatively small error. There are,
however, some particularly dangerous perturbations, which may propagate into the final catalog with
considerable magnification. Such harmful systematic effects should be identified and mitigated, if
possible. This requires numerous mission simulations with different initial data, which may not be
feasible for the extremely computer-intensive solutions for millions of individual grid objects. The
proposed technique is fast enough to be used for massive simulations of slowly varying systematic perturbations, when the emerging astrometric error is confined to the lower degrees
of spherical harmonics.

\begin{figure}[here!]
\epsscale{1.15}
\plottwo{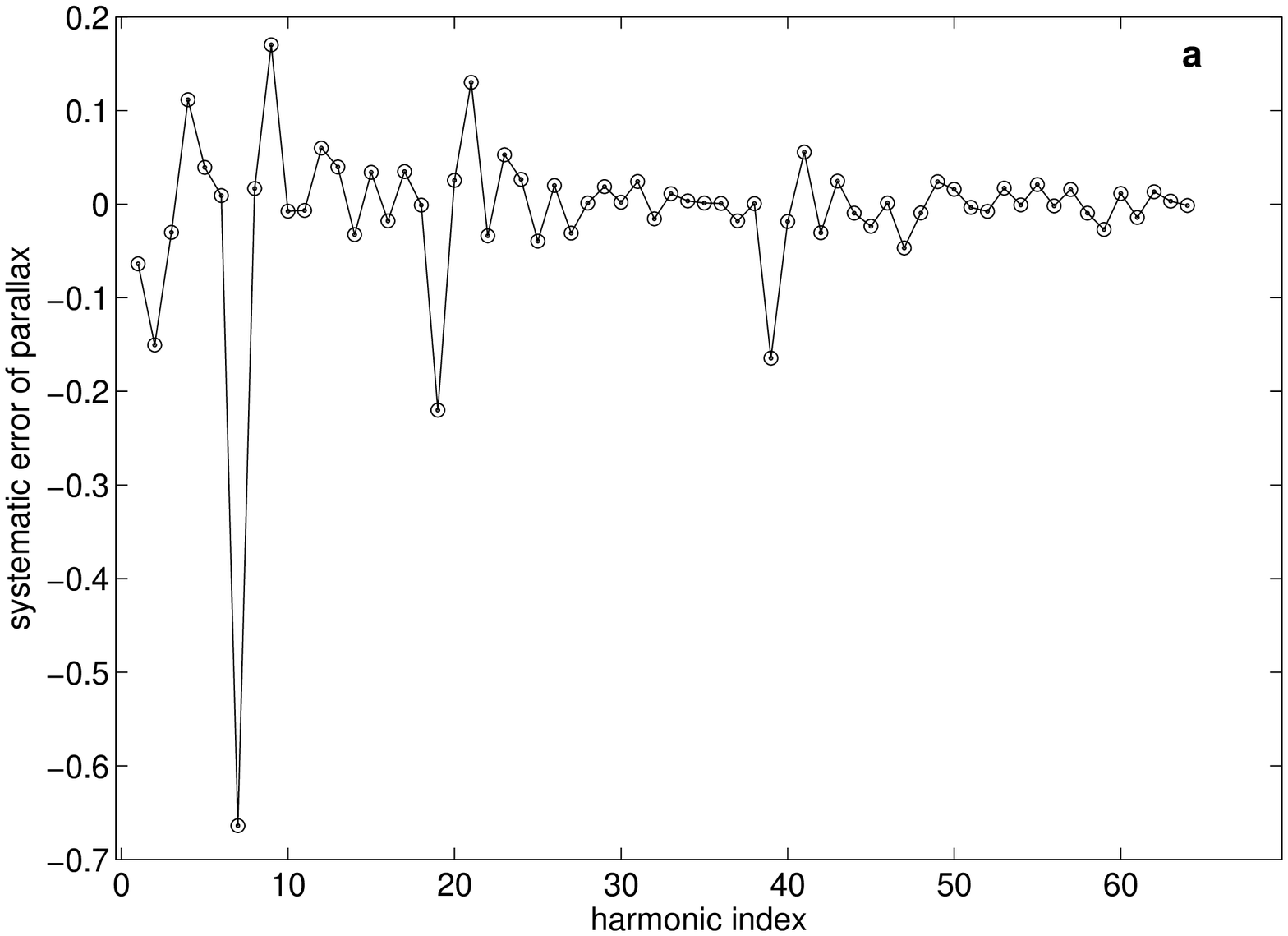}{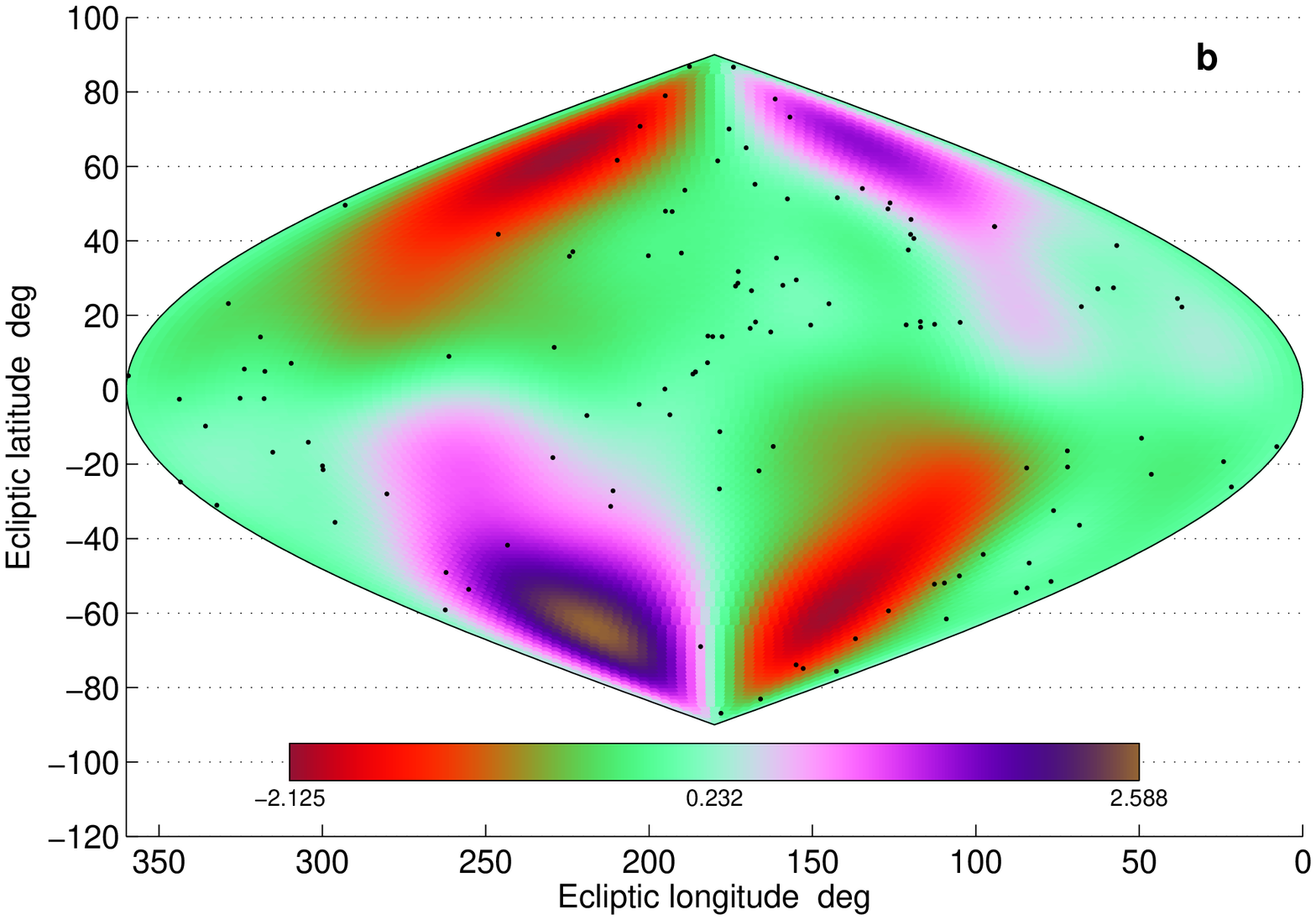}
\caption{Systematic error of JMAPS parallax resulting from a constant perturbation of the differential
scale by 1 mas at the edge of the field of view: a) in the coefficients of the first 64 spherical harmonics;
b) on the sky. All values are in mas.}
\label{syst.fig}
\end{figure}

Figure \ref{syst.fig} shows the results of a specific simulation for JMAPS, where a perturbation in
the field-dependent calibration parameter $Z_2$ was injected in the observational data, but not fitted out
in the global solution. The term $Z_2$ (second Zernike polynomial) corresponds to the differential scale of the instrument. The magnitude of the perturbation was normalized to 1 mas at the edge of the field of view. 
The simulated observations ($X$ and $Y$ measurements) were free of random noise, to see more clearly
the emerging pattern of the correlated error. The absolute error in the coefficients of 64 lower-order
spherical harmonics is depicted in Fig. \ref{syst.fig}a. The spectrum of the error is dominated by
the harmonic number 7 (which is $U_2^0$), followed by harmonic 19 ($U_4^{-2}$) and so on. The total absolute
error of parallax at a given point is the sum of all spherical harmonic errors. The total error in the
first 64 harmonics is depicted in Fig. \ref{syst.fig}b. It shows that the simulated perturbation
is one of the harmful errors for JMAPS, because it compounds to a perturbation of up to 2.6 mas
in some parts of the sky, which is larger than the initial magnitude. Clearly, the distribution of
constraining quasars, shown with black dots, plays a major role in the propagation of this systematic error,
which compounds to larger values in the areas where the quasars are few. If the calibration term $Z_2$ is
included in the set of fitting parameters in the global solution, the emerging error is zero in
the absence of random noise.

\section{Conclusions}
We developed a method to investigate the properties of very large astrometric solutions, which involve
unknown parameters for millions of celestial sources, as well as millions of nuisance unknowns. The method
is based on a stepwise elimination of the attitude and calibration unknowns and the replacement of
individual astrometric corrections with their expansions in orthogonal spherical functions of celestial
coordinates. This approach works well for the JMAPS and SIM
missions and could potentially be useful for Hipparcos
and Gaia. However, demonstrating the applicability of the method to Hipparcos-like missions
would require considerable adjustments, mostly related to the dynamic character of attitude
parameters, which is beyond the scope of this paper. In particular, fitting a set of dynamic parameters
for each extended interval of uninterrupted rotation may render the proposed technique
of QR-elimination of the attitude unknowns impractical. An additional complication arises for Gaia,
where each of the multiple CCDs in the focal plane requires a separate set of calibration
parameters. The pointing, or step-stare mode of operation
of JMAPS makes it best-suited for the proposed global solution technique with block-wise elimination of
attitude and calibration parameters, so that complete analysis for realistic sky coverages and observing
schedules can be performed for billions of condition equations. 
The propagation of zonal and correlated errors of both accidental and systematic origin can be
successfully computed using this method. When the number of expansion terms is appropriately small,
full mission solutions can be obtained using 
regular computers within a few hours with more than a hundred million
unknowns. Some applications of the spherical harmonic solution to the JMAPS mission are described.

\begin{acknowledgements}

\end{acknowledgements}
\end{document}